# CONSTRAINTS ON GRAVITATIONAL DIPOLE RADIATION FROM PULSARS

C Sivaram

Indian Institute of Astrophysics, Bangalore

**Abstract:** Recent suggestions for a modification of general relativity to provide an alternative approach to gravity in connection with the dark energy (matter) problem imply a long range vector component of the gravitational field. This could lead to emission of gravitational dipole emission from objects such as pulsars. Stringent observational limits on period changes of binary and millisecond pulsars and their consistency with general relativity impose severe limits on couplings of such forces. These bounds are tighter than those implied by lunar laser ranging experiments.

The possibility of the existence of long range vector forces in addition to electromagnetism has been raised now and then; starting from Lee and Yang who envisaged such a force coupling to baryon number.[1] More recently there have been several experiments based on various theoretical considerations to search for intermediate range vector forces coupling to isospin or hypercharge.[2] These forces would violate the Einstein equivalence principle and tight limits on their coupling relative to gravity are implied by such experiments.[3] Deviations from Newton's inverse square law over short sub-millimetre distance scales are now being sought mainly motivated by ideas of large extra dimensions, forces due to dark matter particles like axions and Casimir type effects, due to dark energy, etc.[4, 5, 6]

More recently in an attempt to give a relativistic basis for MOND[7], which has had a good amount of phenomenonological success in describing galaxy flat rotation curves without involving dark matter, a tensor-vector-scalar (TeVeS) theory has been proposed as an alternative to Einstein's general relativity.[8, 9] Cosmological tests for such modified theories have been proposed.[10] These types of theories also involve a long range vector component of gravity. The laboratory tests of additional long range vector forces are essentially static experiments, mainly looking for a violation of the equivalence principle through a Eotvos type torsion balance or searching for deviations from Newtonian inverse square law.

However as pointed out in ref [11 and 12] such additional vector forces would have radiative effects in competition with gravitational radiation from astronomical sources, especially from binary systems consisting of compact objects. Indeed constraints on the strength of such forces were put from the known accuracy to which general relativity has been tested for the binary pulsar,[11, 13] as regards the changing separation (changing period) between the components.

In analogy with electromagnetism, we would expect additional vector forces to give rise to electric and magnetic dipole radiation which would cause additional effects in slowing down compact objects like neutron stars. As pulsar spins and periods have been determined to high level of accuracy, this could be used to stringently constrain the strength of such forces. The binary (Hulse- Taylor) pulsar, 1913-16 has now been monitored for over thirty years. The change in orbital period, due to emission of gravitational quadrupole radiation as given by general relativity, is now found to agree with Einstein's theory to within 0.1 percent.

If $M$ be the total mass, $R$ the separation and W the angular frequency of revolution, then the quadrupole energy loss due to gravitational radiation is:

$$\dot{E}_Q \approx \frac{32G}{5c^5} M^2 R^4 \Omega^6 \qquad \ldots (1)$$

Vector forces would in addition give rise to dipole radiation. If there is a gravitational dipole of relative strength $\alpha$ as compared to gravity, ($\alpha$ << 1), then the emission rate of gravitational dipole radiation is given by:

$$\dot{E}_D \approx \frac{2}{3} \alpha \frac{G}{c^3} M^2 R^2 \Omega^4 \qquad \ldots (2)$$

So for typical value of $M$ (1.4 $M_{sun}$), $R$ (~20 km), etc. (eg. binary pulsar) and the fact that there is agreement to 0.1 percent (with quadrupole radiation) implies constraint on the strength, i.e. $\alpha$ of

$$\alpha < \frac{v^2}{c^2} 10^{-3} < 10^{-10} \qquad \ldots (3)$$

The (orbital) velocity $v$ for the binary pulsar is ~300 km/s. Similar constraints would result from the 2.4 hour period binary pulsar recently discovered. The very precise slow down of the millisecond pulsar, i.e. given by $\frac{\dot{P}}{P} \sim 10^{-19} s^{-1}$ implies:

$$\alpha \frac{2G}{c3^3} M^2 R^2 \Omega^4 < B^2 R^6 \Omega^4 \big/ 6c^3 \qquad \ldots (4)$$

$B$ is typically $10^8$ G, for millisecond pulsars. Thus the millisecond pulsar slow down implies:

$$\alpha < 10^{-20} \qquad \ldots (5)$$

I.e. any dipole (vector) component of the gravitational field is <u>$<10^{-20}$ of the Newtonian field strength</u>. For a collapsing star radiating energy, similar limits can be put through:

$$\dot{E}_D \approx \alpha \frac{G}{c^3} P \dot{M} \qquad \ldots (6)$$

Again the merger time due to gravitational radiation is $\sim 5R_0^4 c^5 \big/ 256 M^2 \mu G^3$ where $\mu$ is the reduced mass. This gives a time scale of a billion years for the H-T binary pulsar and a few million years for the 2.4 hour binary pulsar. On the other hand, for dipole radiation, the merger time is proportional to $R_0^3$ (not $R_0^4$!). This could have testable implications for widely separated binaries (like black hole binaries, such as OJ287).[14]

Dipole radiation would cause a rate of change in position of the centre of mass of the system (eg. $\dot{r}_{cms} \sim \dot{E}_D / \mu$, etc[13]). The 'magnetic' part of $\dot{E}_D$ (as given by (2)) would be smaller by a factor of ($v/c$) (i.e. $<10^{-3}$ for the binary pulsars but would however be 0.1 for the millisecond pulsar, see ref. [13]). Again the 'chirping' exponent in analysing the increasing frequency as the binaries spiral is different for dipole. For energy loss, proportional to $r^{-k}$, the chirp frequency is $(t_0 - t)^{-3/2(k-1)}$. For dipole, $k = 3$ and this is $-1/2$.

A more detailed discussion of how such radiation can affect gravitational wave detectors (this was discussed in connection with intermediate range forces) is given in ref. [12] and the implication for dark matter in ref. [13].

It must be pointed that lunar laser ranging (LLR) currently provides the best constraint on deviation from Newtonian force law as $10^{-10}$ times the strength of gravity (at $10^8$ metre scale).[15] This is comparable to the above limits. The millisecond pulsar gives a much tighter limit, ten orders more stringent.

Also P and T violating gravitational interactions have been postulated with couplings of the form: $R_{\alpha\beta\gamma\delta} R^{*\alpha\beta\gamma\delta}$ ($R^*$ the dual curvature) similar to QCD T violating terms $\sim F_{\alpha\mu\nu} F^{*\alpha\mu\nu}$. Other terms involve derivative curvature couplings like $\phi\partial_\mu R$ etc have been proposed. Such interactions have been invoked in connection with baryogenesis in the early universe and would lead to gravitational dipole moments.[16]

The above astronomical limits would tightly constrain the coupling of such terms in the gravitational action.